# Direct observation of the ice rule in artificial kagome spin ice

Yi Qi[1], T. Brintlinger[1,2], and John Cumings[1*]

[1]Department of Materials Science and Engineering

[2]Center for Nanophysics and Advanced Materials

University of Maryland, College Park, 20742

[*]Corresponding author. email:cumings@umd.edu

**Abstract:**

Recently, significant interest has emerged in fabricated systems that mimic the behavior of geometrically-frustrated materials. We present the full realization of such an artificial spin ice system on a two-dimensional kagome lattice and demonstrate rigid adherence to the local ice rule by directly counting individual pseudo-spins. The resulting spin configurations show not only local ice rules and long-range disorder, but also correlations consistent with spin ice Monte Carlo calculations. Our results suggest that dipolar corrections are significant in this system, as in pyrochlore spin ice, and they open a door to further studies of frustration in general.

75.75.+a, 75.50.Lk

**Text:**

Geometrical frustration is known to significantly modify the properties of many materials. Pyrochlore spin ice and hexagonal water ice are canonical systems[1, 2] that show the effects of frustration in both heat capacity[3, 4] and dynamical response[5, 6], and frustration also influences the mechanical response of water ice[7], with geologically significant implications. In both instances, microscopic ordering principles on the lattice lead to a macroscopic degeneracy of configurations. This degeneracy may also be modified or lifted by lattice imperfections, as in the case of KOH-doped water, where a first-order transition to an ordered ground state emerges[8]. Unfortunately, these effects are difficult to model or predict, because existing experimental techniques cannot directly observe the local ordering, near lattice defects or otherwise. To address this long outstanding problem, recent interest [9-14] has focused on fabricating systems that allow the effects of frustration to be physically modeled and the resulting local configurations to be directly observed.

A prominent example of the approach is the recent work of Wang et al., who demonstrate a lithographic pattern of nanoscale islands of magnetic material that may behave as a 2D analog of pyrochlore spin ice, dubbed "artificial spin ice"[9]. However, the realization does not obey its corresponding "two-in two-out" ice rule, which would lead to frustration. Instead, the system only exhibits a statistical preference for the ice-rule-obeying configurations among a disordered distribution that includes all possible configurations, some expressly forbidden by the ice rule. This is unanticipated for a model system where the corresponding ice-rule temperature is expected to be on the order of $10^5$ K. Furthermore, a theory developed in Ref. 12 to describe the system invokes only a short-range vertex interaction; an interesting and potentially significant component of the pyrochlore spin ice model is the long-range dipolar interaction, which might possibly lead to long-range order in the material [15]. We here present a new realization of artificial spin ice that both strictly obeys its local ice rule and also shows the effects of long-range dipolar interactions.

As a starting point for our realization, we use the honeycomb magnetic structure, also proposed by Tanaka et al.[16] In their study, it is demonstrated that the honeycomb magnetic structure can be mapped onto a spin ice system on the kagome lattice, as shown in figure 1. Interestingly, this same lattice has also been studied theoretically using Metropolis Monte Carlo simulations [17]. The kagome lattice is a two-dimensional structure composed of corner-sharing triangles. It is an essential component of the pyrochlore spin ice structure[10, 18-20] and has also been connected with jarosite frustrated magnets[17,21]. Compared to the 2-in-2-out ice rule for the pyrochlore structure, the ice rule here changes to 2-in-1-out or 1-in-2-out for each vertex (see Fig. 1c). In the structure, the magnetization along each connecting element of the honeycomb lattice adopts a single domain, and the domain walls are constrained within the vertices, where micromagnetic energies allow only the ice-rule-obeying configurations. Unfortunately, the study of Tanaka et al. was not able to uncover the local magnetization within each element, as we show below. Additionally, the study does not include an energy-minimizing protocol, exploring the possibility of ice-rule violating vertices among random ensembles, as Wang et al. have done. We here present an "artificial spin ice" approach to the magnetic honeycomb structure that addresses both of these experimental deficiencies.

In Ref. 16, magnetic force microscopy (MFM) is used to image the magnetic structure of the kagome lattice. MFM works by detecting escaped flux from the material, and in these structures it can therefore only yield information about the excess flux at a given interaction vertex. The kagome lattice possesses three magnetic elements per Bravais lattice point, with each element having a two-level degree of freedom. However, MFM can only capture two-level information at the interaction vertices, which number two per Bravais lattice point. For a lattice with n Bravais lattice sites, MFM results may express up to $2^{2n}$ unique states, whereas the lattice can exhibit on the order of $2^{3n}$ (both with small corrections for the ice rule).

To demonstrate this deficiency, we use the data and model presented in Fig. 2 of Ref. 16. This figure contains MFM data as Fig 2a and a model of the magnetic moment orientations as Fig 2b, and both are reproduced here as Figs. 2a and 2b, respectively. To explicitly demonstrate the under-defined nature of moment configurations constructed from MFM on interacting vertices, one can construct a new model of magnetic orientations by selecting from a given moment map any head-to-tail chain of elements and reversing the entire chain. Two possible examples are shown here in Figs. 2c and 2d, and we estimate that there are on the order of $2^{12}$ other such configurations, only one of which reflects the actual unknown configuration of the system. This uncertainty also makes second- and third-nearest neighbor correlations impossible to deduce. What is needed to adapt the honeycomb network into a full-fledged physical model of kagome spin ice is an imaging technique that directly and unambiguously records the internal magnetic flux of the wire elements. This can be achieved by Lorentz-mode transmission electron microscopy, as we demonstrate below.

Our realization of the kagome structure is fabricated from permalloy ($Ni_{80}Fe_{20}$) using conventional electron-beam lithography, followed by metal deposition and lift-off. Fig. 2a shows a transmission electron microscope (TEM) image of our structure. The lines of the honeycomb are 500nm long, 110nm wide and 23nm thick. At this scale, micromagnetic simulations[22] indicate that the connecting elements are magnetized along their axis and act as macroscopic Ising spins with energy differences among the different configurations that support the ice rule assumption[23]. With strong analogies to real spin ice, these simulations show that 85% of this nearest neighbor energy difference comes from dipolar field, with the remaining 15% coming from exchange energy due to the domain walls at the vertices. The total number of elements in our realization is 12,864, large enough for ensemble results comparable with Monte Carlo simulations [17].

To determine the directions of the single-domain elements, we employ a TEM operating in Lorentz imaging mode, which is traditionally used to detect domain structures of magnetic materials[24, 25]. To simulate the contrast of single-domain needle-shaped elements, we use a standard contrast transfer function[26]. Fig. 2c shows that images of the spin elements have over-focus Lorentz contrast featuring a dark edge and a bright edge, depending on the magnetization direction. Simply, this can be explained by Lorentz-force deflection when the electron beam passes through a magnetic element. Fig. 2b shows a Lorentz-mode image corresponding to Fig. 2a, and we can see the elements have varied contrast because of their varied magnetization directions. Using a right-hand rule, we uniquely specify a direction for each element, as shown by the colored arrows. We verify the magnetic origin of the contrast both by through-focus imaging and by in-situ field reversal.

To coerce the structure toward its magnetic ground state, we demagnetize the sample using a decreasing and rotating magnetic field prior to imaging, following the procedure of Wang et al.[27] The demagnetizing process introduces varied vertex configurations into the lattice. Fig. 3a shows a spin map of part of the kagome lattice after the demagnetization process, where we utilize a color wheel to represent different spin directions. Consequently, neighboring elements with close colors have a head-to-tail low-energy configuration, while those with opposing colors have a head-to-head or tail-to-tail high-energy configuration. A first glance reveals that the spins are quite disordered in long range, a signature found in most frustrated systems.

For detailed statistical studies of the spin distributions, we count the elements using a numerical method, labeling spins pointing to one of the two Ising directions as $s_i = 1$, and the opposite directions as $s_i = -1$. The net magnetization is then defined as $m = \langle s_i \rangle$ for each of the three sub-lattices of spins. The demagnetization process typically achieves $|m|$ in the range of 0.03-0.14. The distribution of vertex types is plotted in Fig. 3b and varies among the six ice-rule vertex types from 9.8% – 24.8% [28]. We find that all vertices fall into the six low-energy configurations, and there are no 3-in or 3-out high-energy states. Therefore, *every* vertex satisfies the ice rule. This is the first experimental demonstration of rigid adherence to the ice rule in any frustrated system (real or artificial), where each vertex is determined by explicitly counting its local spins.

As expected for ice-rule-governed interactions, our kagome lattice can reach a state with a large degree of disorder and a small net magnetization. Thus, we have an ideal system for calculating the intrinsic correlations defined by lattice geometry and magnetic interactions. Based on the fact that we observe many vertex types and the specific configuration varies from run to run, the correlation calculated would be expected to be close to its intrinsic value according to statistical theory. The correlation between spins $i$ and $j$ is defined as $c_{ij} = 1$ when $\vec{s}_i \bullet \vec{s}_j$ is positive, otherwise $c_{ij} = -1$. Different types of correlations may be calculated based on their relative position, as shown in Fig. 1b. The correlation coefficient is calculated as the average value for each type of such pairs, e.g., $C_{\alpha\beta} = <c_{ij}>|(ij \in \alpha\beta)$. This is mathematically equivalent to the correlations calculated in [17].

The correlation coefficients are summarized in table I and are compared with Monte Carlo simulation results based on a kagome spin ice model using only nearest-neighbor interactions[17]. We note substantial consistency between our results and the model simulation. Specifically, $C_{\alpha\beta} = 1/3$ indicates that *all* vertices obey the ice rule. Each of the other pairwise correlations shows ferromagnetic (positive) or antiferromagnetic (negative) values, agreeing in sign and relative magnitude with Monte Carlo simulations. However, we note our measured higher-order correlations have reproducibly larger absolute values than predicted by Monte Carlo with only nearest-neighbor interactions. As is shown in Table I, the relative dipole energies, calculated using simple magnetostatics for each configuration, agree in sign with the correlation values. This strongly suggests that dipolar interactions play a significant role in the ordering of our model spin ice, as is the case for real spin ice. These long-range interactions generally increase the ordering in spin ice, decreasing the degeneracy of the ground state manifold[15, 29].

We again emphasize that the ice rule is strictly obeyed for the kagome ice system we study, with no instances of non-ice-rule vertices, in contrast to results reported for a square lattice[9]. One possible reason for this is the relatively strong interaction between nearest-neighbors in our connected lattice, including both exchange and dipolar energies and thus making the 3-in or 3-out configuration highly unfavorable. Another reason that no 3-in or 3-out configurations are found might be explained as follows: in the annealing process, changing from a 3-in or 3-out high energy state to a 2-in-1-out or 1-in-2-out only requires one spin to flip and thus proceeds readily, allowing the system to approach an energy minimum. On the other hand, in the square lattice, the analogous process would generally require chain- or loop-flipping with a low probability. In this sense, the kagome lattice we use in the present study is likely "more ergodic" than the square lattice, which explores a demonstrably limited range of parameter space[12].

These results demonstrate that the magnetic honeycomb structure is an ideal artificial spin ice system for studying the effects of frustration. Its simplicity and ease of fabrication make it a robust platform for studying the possible influence of lattice imperfections in geometrically frustrated physical systems. Additionally, it achieves this without need for mathematical approximations or lengthy computations[30] and without the trial-and-error typically associated with materials discovery. As demonstrated by the relatively good agreement between our correlations and the results of Monte Carlo simulations, the demagnetization process we employ might also serve in a more general sense as an efficient proxy for other computer models that search for optimal solutions in configuration space. With appropriate modifications, the artificial spin ice approach may open a door to solving other optimization problems as well.

The authors acknowledge useful conversations with R. F. Wang, P. Schiffer, O. Tchernyshyov, M. S. Fuhrer, and T. Einstein. This research was supported by funding from the Minta Martin Foundation, the National Science Foundation under grant DMR-075368, and the NSF-MRSEC at the University of Maryland, grant DMR-0520471.

**Figure 1.** (color online) (a) A sketch of the kagome spin ice lattice showing 30 spins. The two sublattices on which interaction vertices can occur are labeled ● and ○. (b) The honeycomb structure formed by connecting the spins of the kagome lattice. Each bar element represents a spin magnetic moment oriented along the bar axis. The Greek symbols label spins for later use in correlation calculations. (c) The spin configurations possible at a single vertex. Spin configuration that obey the ice rule produce a net magnetic moment at each vertex, which we use to label the allowed spin configurations. The two configurations that produce no net magnetic moment (3-in and 3-out) are not energetically favorable.

**Figure 2.** (color online) a) MFM data presented by Tanaka et al. [16] b) A spin configuration proposed therein to describe the data. c) and d) Valid alternative configurations obtained by reversing chains of elements, shown in different colors. In the general case, these chains include closed loops, which can be reversed between clockwise and counter-clockwise.

**Figure 3.** (color online) (a) An in-focus TEM image of our fabricated kagome structure, scale bar: 1 μm. Inset: a design image of the entire lattice, scale bar: 10 μm (the individual elements cannot be seen at this scale). (b) A TEM image of the same kagome structure with Lorentz contrast. (c) A Lorentz TEM simulation using a contrast transfer function reveals the single domain magnetic moment direction

based on the dark-bright edge contrast; using this, six spins in (b) are labeled with their directions. The two circles in (b) indicate CW and CCW closed loops.

Figure 4. (color online) (a) A region of the spin map from a demagnetized kagome lattice sample. The spin directions are disordered in long-range, with a small net magnetization, yet locally there are some ordered chains and loops. (b) The vertex-type distributions; three demagnetization data sets are shown with differently shaded bars. The bar labels are from Fig. 1c. The percentage of each type of vertex ranges from 9.8%-24% and varies from run to run.

Table I. Correlation coefficients calculated from a demagnetized sample. The results are shown in mean and standard deviation taken from three demagnetization runs. The model values are from [17]. $\Delta E_{dipole}$ gives the energy difference between aligned and anti-aligned spin pairs, normalized to the nearest neighbor value.

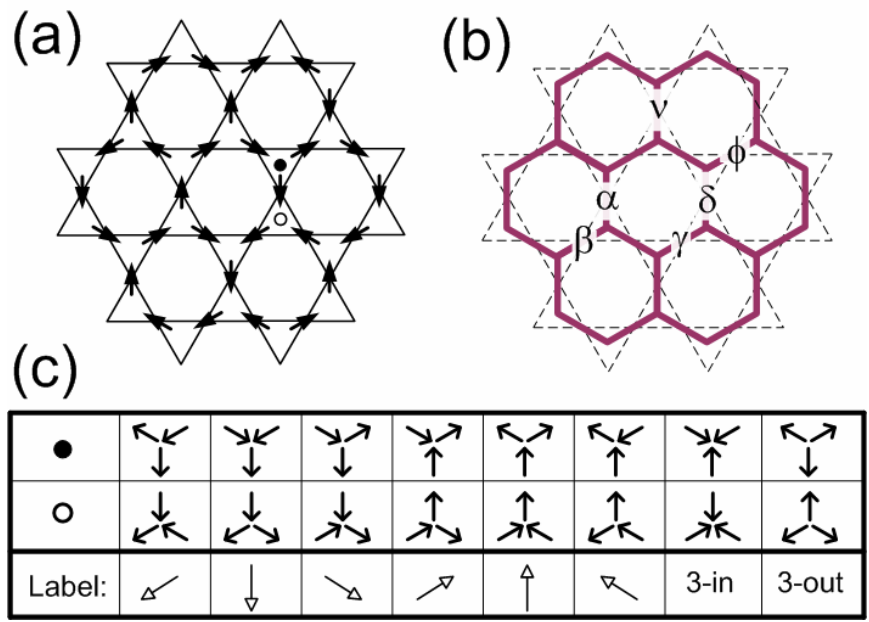

**Figure 1.**

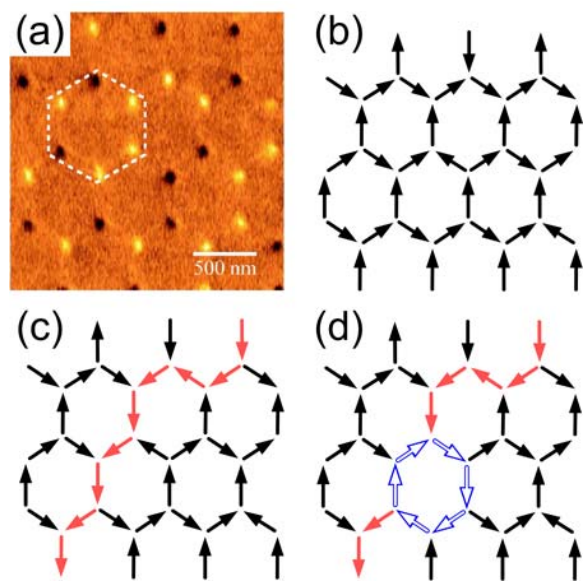

**Figure 2.**

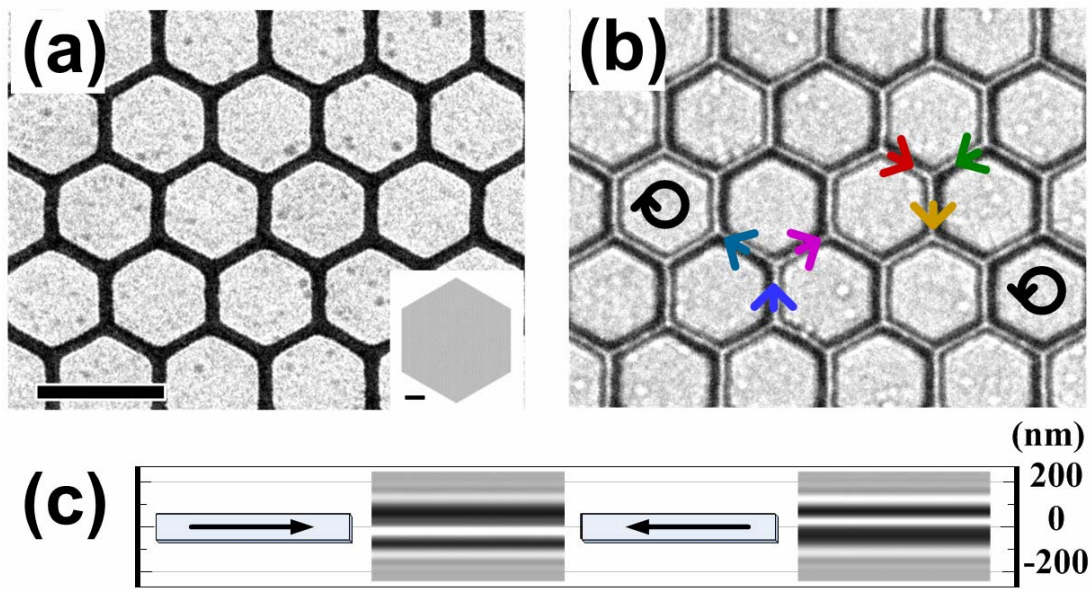

**Figure 3.**

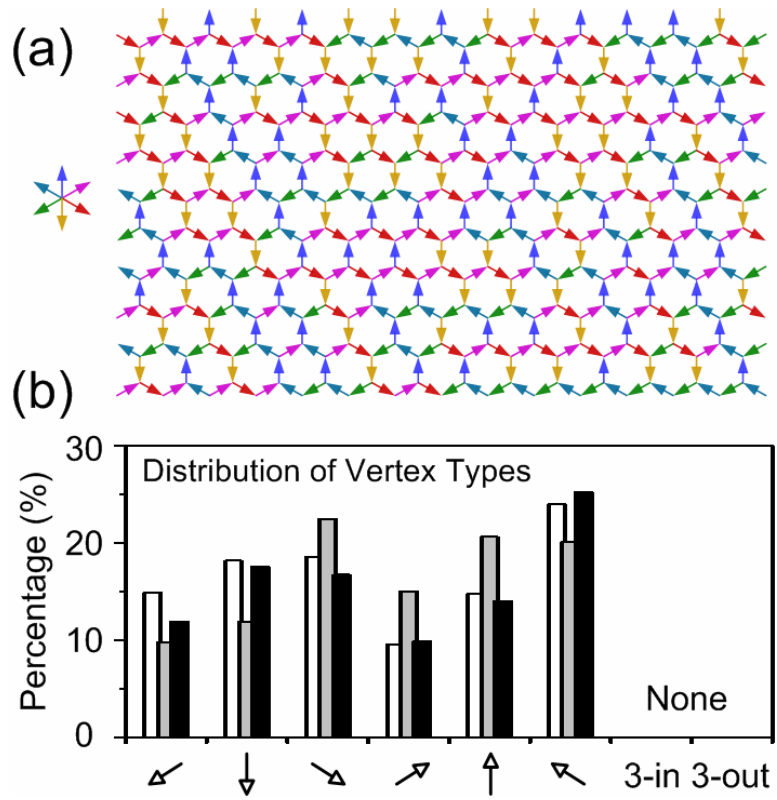

**Figure 4.**

|  | Data | Model | $\Delta E_{dipole}$ |
|---|---|---|---|
| $C_{\alpha\beta}$ | 0.333 | 0.333 | *1.0* |
| $C_{\alpha\gamma}$ | -0.158±0.008 | -0.118 | *-0.137* |
| $C_{\alpha\nu}$ | 0.165±0.013 | 0.101 | *0.089* |
| $C_{\alpha\delta}$ | -0.130±0.015 | -0.072 | *-0.070* |
| $C_{\beta\phi}$ | 0.057±0.007 | 0.007 | *0.082* |

**Table I.**